\documentclass[conference]{IEEEtran}
\usepackage{cite}
\usepackage{amsmath,amssymb,amsfonts}
\usepackage{algorithmic}
\usepackage{graphicx}
\usepackage{textcomp}
\usepackage{xcolor}
\usepackage{epstopdf} 
\usepackage{url} 
\usepackage{mathtools}
\usepackage{gensymb}
\usepackage[nonumberlist]{glossaries}
\newacronym{puf}{PUF}{Physical Unclonable Function}
\newacronym{ppuf}{PPUF}{Public Physical Unclonable Function}
\newacronym{iot}{IoT}{Internet of Things}
\newacronym{cots}{COTS}{commercial off-the-shelf}
\newacronym{adc}{ADC}{analog-to-digital converter}
\newacronym{dac}{DAC}{digital-to-analog converter}
\newacronym{ac}{AC}{alternating current}
\newacronym{rms}{RMS}{root mean square}
\newacronym{rmse}{RMSE}{root mean square error}
\newacronym{soc}{SoC}{System-on-a-chip}
\newacronym{uno}{UNO}{unique object}
\newacronym{mcu}{MCU}{Microcontroller}
\newacronym{mpu}{MPU}{Microprocessor}
\newacronym{chap}{CHAP}{Challenge-Handshake Authentication Protocol}
\newacronym{scram}{SCRAM}{Salted Challenge Response Authentication Mechanism}
\newacronym{is}{IS}{Impedance Spectroscopy}
\newacronym{dft}{DFT}{Discrete Fourier Transform}
\newacronym{lpwan}{LPWAN}{Low-Power Wide Area Network}

\ifCLASSOPTIONcompsoc
    \usepackage[caption=false, font=normalsize, labelfont=sf, textfont=sf]{subfig}
\else
\usepackage[caption=false, font=footnotesize]{subfig}
\fi

\def\BibTeX{{\rm B\kern-.05em{\sc i\kern-.025em b}\kern-.08em
    T\kern-.1667em\lower.7ex\hbox{E}\kern-.125emX}}
\begin{document}

\title{Fingerprinting Analog IoT Sensors for\\Secret-Free Authentication}

\author{
\IEEEauthorblockN{
  Felix Lorenz\IEEEauthorrefmark{1}, Lauritz Thamsen\IEEEauthorrefmark{1}, Andreas Wilke\IEEEauthorrefmark{2}, Ilja Behnke\IEEEauthorrefmark{1}, Jens Waldmüller-Littke\IEEEauthorrefmark{2},\\ Ilya Komarov\IEEEauthorrefmark{2}, Odej Kao\IEEEauthorrefmark{1}, and Manfred Paeschke\IEEEauthorrefmark{2}
\IEEEauthorblockA{
  \IEEEauthorrefmark{1}Technische Universit\"at Berlin, Germany, \{firstname.lastname\}@tu-berlin.de\\
}
\IEEEauthorblockA{
  \IEEEauthorrefmark{2}Bundesdruckerei GmbH, Germany, andreas.wilke@bdr.de
}
}
}

\maketitle

\begin{abstract}
Especially in context of critical urban infrastructures, trust in IoT data is of utmost importance.
While most technology stacks provide means for authentication and encryption of device-to-cloud traffic, there are currently no mechanisms to rule out physical tampering with an IoT device's sensors.
Addressing this gap, we introduce a new method for extracting a hardware fingerprint of an IoT sensor which can be used for secret-free authentication.
By comparing the fingerprint against reference measurements recorded prior to deployment, we can tell whether the sensing hardware connected to the IoT device has been changed by environmental effects or with malicious intent.
Our approach exploits the characteristic behavior of analog circuits, which is revealed by applying a fixed-frequency alternating current to the sensor, while recording its output voltage.
To demonstrate the general feasibility of our method, we apply it to four commercially available temperature sensors using laboratory equipment and evaluate the accuracy.
The results indicate that with a sensible configuration of the two hyperparameters we can identify individual sensors with high probability, using only a few recordings from the target device.
\end{abstract}


\section{Introduction}\label{sec:introduction}
\gls{iot} technology provides a promising path towards improved monitoring and control of critical urban infrastructures such as transport systems, water networks, and telemedicine systems~\cite{rashid2016}.
With the stringent security requirements of these application domains, reliable means to establish trust in the sensor data streams is of absolute necessity.
Established cryptographic protocols can be used to protect the data from malicious third parties while it traverses the fog.
However, in many cases ~\gls{iot} nodes are situated in physically vulnerable, publicly accessible spots, where attackers can gain access to a device and tamper with its peripherals.
In other words, even if a device's identity has been established and is therefore trustworthy, how can we verify that the signals coming from its sensors are trustworthy as well?
In the literature, this problem is called~\emph{sensor-crypto separation}~\cite{rosenfeld2010} and generally refers to the possibility of intercepting signals between an analog sensing device and the IC processing the results. 

\begin{figure}[htb]
  \centering
  \includegraphics[width=0.98\columnwidth]{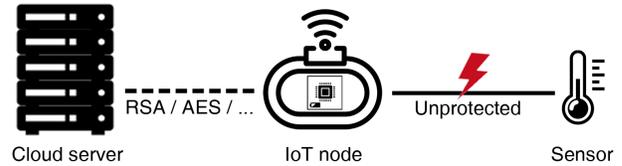}
  \caption{The sensor-crypto separation problem: While traffic between the IoT device and the cloud is encrypted, an attacker can manipulate the measurements while they traverse the wire between sensor and device.}\label{fig:sensorcrypto}
\end{figure}

In this paper, we address the problem of sensor-crypto separation with a novel method for fingerprinting the sensors connected to an~\gls{iot} node.
Our method is based on the application of alternating currents to the sensor inputs and reading the sensor outputs.
Before deployment of a sensor-equipped device, a broad frequency spectrum is scanned and the results are stored in a database.
Later, when a device wants to prove that its sensing hardware has not changed, it transmits the responses to a small set of probe frequencies.
The remote party can then compare the received responses with the ones in the database and, based on their deviation, decide whether the configuration is likely to have changed in-between deployment and the current response or not.
Identification granularity can be divided into two categories:
\begin{itemize}
    \item \emph{Inter-device} identification distinguishes between different sensor models.
    \item \emph{Intra-device} identification distinguishes between different instances of the same model.
\end{itemize}
Of course, the latter is more difficult than the former, and any method that can reliably distinguish between sensor instances can also distinguish models.
Using a set of~\gls{cots}~\gls{iot} sensors and standard laboratory equipment, we demonstrate the general feasibility of our approach and discuss its strengths and limitations.
We conclude that our method is powerful enough to perform inter- and intra-device identification with bounded error.\\



The rest of this paper is structured as follows:
First, we introduce a few relevant concepts in Section~\ref{sec:background} and discuss related publications in Section~\ref{sec:relatedwork}.
Then, we introduce our method in Section~\ref{sec:method} and evaluate it for several commercial IoT temperature sensors using precision measurement equipment in Section~\ref{sec:evaluation}.
Finally, in Section~\ref{sec:discussion} we provide a discussion of the results, before we conclude the paper and highlight our plans for further improvements in Section~\ref{sec:conclusion}.

\section{Background}\label{sec:background}

The fingerprinting technique discussed in this paper relates to some concepts from the security domain, which are briefly introduced below.

\paragraph{Challenge-response authentication}
Some of the fingerprinting techniques discussed in Section~\ref{sec:relatedwork} are based on the~\emph{challenge-response principle}:
To authenticate a device, the remote party sends a question (``challenge'') for which the local party must produce a valid answer (``response'').
Before deployment, the system is~\emph{bootstrapped}, which means that for all possible challenges, the corresponding responses of the target system are recorded and stored.
Later, checking the validity of any single response corresponds to a simple comparison with the database of challenge-response pairs.
For reliable identification the device should consistently produce the same response to a given challenge and no two devices must produce identical responses to all available challenges). 
Hardware implementations of the challenge-response principle predominantly rely on~\emph{process variation} during manufacturing to ensure uniqueness.

\paragraph{Secret-free authentication}
Cryptographic schemes often include the establishment of a~\emph{shared secret} between the involved parties that is used for authentication and encryption.
Naturally, a wide range of attacks have been devised that focus on extracting the secret, e.g. using malware or through some physical side channel.
To eliminate this class of attacks, researchers are trying to develop~\emph{secret-free} schemes, where everything about both ends of a conversation is known and yet an attacker has no way of impersonating one or the other party~\cite{ruhrmair2019}.
Current examples of secret-free authentication are presented in the next section.

\section{Related Work}\label{sec:relatedwork}

Many~\gls{iot} solutions rely on~\gls{lpwan} technology for communication and the popular implementations support block cypher encryption between devices as well as application server.
Therefore, the research community has primarily focused on identifying the type of node connected to a network based on the observed network traffic~\cite{bezawada2018,marchal2019,nguyen2019}.
To the best of our knowledge, no works have been published that go beyond the end device to determine if its sensor configuration has changed.
In the following sections, we revise fingerprinting methods that rely on similar physical characteristics as the one described in this paper, despite the fact that they were not specifically developed for~\gls{iot} applications.

\subsection{(Public) Physical Unclonable Functions}\label{sec:ppuf}

\glspl{puf} are special circuits designed to specify a device with a unique and unclonable identity for authentication.
The circuit takes a bit pattern as input and deterministically produces another bit pattern as output.
Process variations during manufacturing make the mapping between inputs and outputs unique and prevents an attacker vom physically duplicating the circuit.
In the recent past,~\glspl{puf} have fallen slightly out of favor due to their susceptibility to a class of attacks which use Machine Learning to simulate the~\gls{puf} behavior~\cite{ruhrmair2010}.

This weakness was later addressed through the introduction of~\glspl{ppuf}~\cite{beckmann2009}.
Everything about a~\gls{ppuf}'s circuit is public knowledge and yet the system is designed in a way that an attacker cannot simulate the device efficiently, i.e. simulating a response takes significantly longer than just physically producing it with the original circuit.
By verifying the timing of the response in addition to its correctness, the remote party can verify that it is dealing with the real device and not a simulation of it.
The~\gls{puf} principle can be transferred to various kinds of materials that are also subject to process variation, such as CDs and sheets of paper~\cite{ruhrmair2012}.
In this case, the medium is called~\gls{uno} and the fingerprint is revealed using a high-resolution measurement device, such as a microscope.
\glspl{uno} and~\glspl{ppuf} constitute prime examples of secret-free cryptographic systems and are based on the same principle as our method:
To reveal unique properties of physical objects that result from process variations.\\

Both~\gls{puf} and~\gls{ppuf} have been suggested as suitable technologies for authentication protocols in~\gls{iot} ecosystems.
In most cases, the circuit is directly integrated with the~\gls{iot} node or added as a peripheral identification module.
But there have also been efforts to combine a sensing circuit and a fingerprint circuit into a sensor that can do both at the same time: measuring a target observable and identifying itself.
In~\cite{rosenfeld2010} the authors propose to fuse a photosensor with a~\gls{puf} to produce a device that signs each measurement with a response~\emph{before} it leaves the~\gls{puf} circuit. 
Similar schemes have been described for measuring pressure~\cite{tang2016} and voltage~\cite{gao2017}, and capacitance~\cite{karuthedath2018} respectively.
Others go further and use the environment-dependent behavior of a~\gls{puf} directly for sensing~\cite{shimizu2015}, albeit with serious limitations:
While they prove that it is possible to detect faults injected into the devices power supply, it remains an open question whether the idea can be transferred to sensors for other physical quantities.
Unfortunately, all approaches in which a sensor is directly combined with a PUF introduce dedicated circuits, which means they are incompatible with~\gls{cots} sensors currently on the market.
Our method addresses this gap by revealing the uniqueness of a sensor itself instead of relying on the entropy of a~\gls{puf} circuit.


\subsection{Smartphone and Radio Fingerprinting}\label{sec:smartphone}

In the recently emerging field of smartphone fingerprinting, the goal is to identify a smartphone remotely with high probability by revealing the uniqueness of its internal sensors.
One of the earliest methods was developed by Dey and colleagues~\cite{dey2014} and aims to identify a phone via its built-in accelerometer.
Specifically, the challenge is produced in the form of a vibration from the internal mechanical vibrator and consecutively, the response is captured as accelerometer readings.
Inconveniently the success of their method depends on the surface on which the phone is laying.
To overcome that limitation, another approach was soon proposed by Bojinov et al.~\cite{bojinov2014}.
Here, the challenge is emitted through the phone's speakers in the form of an auditory signal and simultaneously recorded as a response using the microphone.
Their method does not require the device to be in a specific physical location but still depends on a relatively noise-free environment in order to work properly.
Finally, integrated schemes were described for combining the fingerprints of multiple sensors for more reliable identification~\cite{lee2015,amerini2017}.
A comprehensive review of the literature on smartphone identification can be found in~\cite{baldini2017} as well as a discussion of relevant attacks in~\cite{ren2019}.

For devices without built-in sensors and actuators, physical-layer identification techniques focus on unique characteristics of the analog radio circuitry in wireless transceivers~\cite{danev2012}.
They are subsumed under the term \emph{RF-DNA} and construct models of the device classes from the observed radio communication using methods like Wavelet transform~\cite{klein2009} or Hilbert Huang transform~\cite{cobb2011}.

\section{Method}\label{sec:method}

The main idea behind our fingerprinting method is to reveal the characteristic properties of a sensor's analog circuit and use them for secret-free authentication:
Due to the resistive, inductive, and capacitive parts, an analog circuit behaves like a harmonic oscillator upon being stimulated with~\gls{ac}. 
The response is determined by the circuit's layout, the attributes of individual electronic components, the frequency of the~\gls{ac} voltage, and environmental conditions such as air temperature or local magnetic field.
Thus, assuming we can control or at least account for the influence of environmental factors, we can distinguish two sensors by comparing their responses to various input frequencies.
The granularity of identification depends on the resolution of the measurement equipment, since the difference in behavior among circuit layouts is orders of magnitude larger than the difference due to process variation.
It should also be noted that the characteristic behavior of the devices should be revealed at the operational limits of input parameters as specified in the datasheets.
E.g. if a sensor's operational voltage range is specified as 3.3V to 5V, we should test around 3.3V since going above 5V could damage the device.
The intuition behind this is, that the circuits are designed with certain in-built tolerances and compensation mechanisms within the supported ranges that would conceal characteristic behavior.

\subsection{Fingerprint Extraction}

\begin{figure}[htb]
  \centering
  \includegraphics[width=0.8\columnwidth]{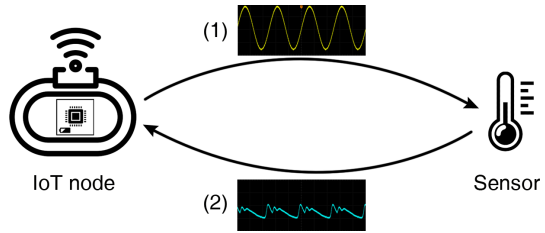}
  \caption{Overview of the proposed challenge-response scheme. (1) AC with fixed frequency is applied to the sensor's inputs. (2) The response is recorded.}\label{fig:method}
\end{figure}

Our approach follows the challenge-response principle: 
The sensor is supplied with an oscillating voltage of a fixed pattern and frequency (\emph{challenge}).
Oscillation range and waveform are selected according to the capabilities of the available hardware. 
Meanwhile, its output voltage $V_{out}$ is sampled for a certain number of steps and used to compute specific parameters of the output waveform (\emph{response}).
In the following, we focus on effective value and variance but briefly discuss other options in Section~\ref{sec:discussion}.
Since the responses of two different sensors can happen to coincide for some frequencies, multiple such challenge-response pairs must be recorded to reduce the number of false positives.
A schematic representation of our method is given in Figure~\ref{fig:method}.\\


\subsection{Authentication scheme}

The fingerprinting method is used as part of an authentication scheme that proceeds similarly to how~\gls{puf}-based authentication works.
Prior to device deployment, a~\emph{bootstrapping} step is performed.
This is essentially a sweep over a predefined frequency band with the responses for each point being recorded.
The resulting set of challenge-response pairs is denoted as~\emph{full fingerprint} of the target device and must be kept secure with the trusted remote party. 
Depending on the~\gls{iot} architecture and network topology, this might be a cloud server or an edge node.
Then, the sensor is connected to an~\gls{iot} node and deployed to its designated location.
To authenticate the sensor in the field, first the corresponding~\gls{iot} node is authenticated using a standard authentication mechanism.
Then, a set of challenge frequencies called~\emph{partial fingerprint} is chosen according to a predefined pattern and the responses are sent to the backend for verification.
Here, we deviate from the usual process of~\gls{puf}-based authentication:
In order to reduce communication overhead and conserve power of the~\gls{iot} node, we do not transmit the challenge from the remote endpoint to the device but instead apply a hashing function locally to the current timestamp to obtain a set of input frequencies. 
If the partial fingerprint deviates from the responses of the full fingerprint by no more than a specified margin, we consider the authentication to be successful and can assume that the measurement originated from the same circuit that was bootstrapped previously.


\section{Preliminary Experiments}\label{sec:evaluation}

\begin{figure}[htb]
  \centering
  \includegraphics[width=0.98\columnwidth]{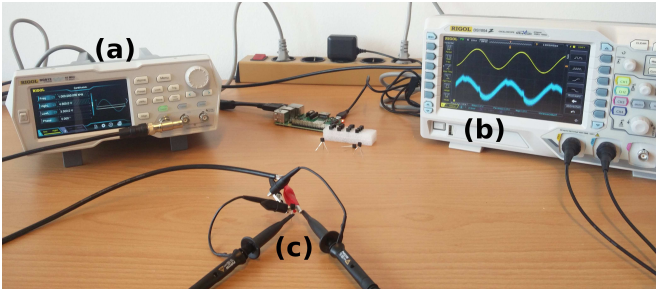}
  \caption{Our experimental setup consisting of a function generator (a), an oscilloscope (b), and the sensor (c).}\label{fig:exp_setup}
\end{figure}

To verify the general feasibility of our approach, we conducted a series of experiments on real~\gls{iot} sensors using laboratory equipment as presented in Figure~\ref{fig:exp_setup}.
Specifically, we use a Rigol DG812 function generator to produce the input oscillations and measure sensor output voltage with a Rigol DS1054Z oscilloscope.
The~\gls{dac} inside the function generator offers a resolution of 16 bits with a sampling rate of 125 MSa/s, whereas the oscilloscope's~\gls{adc} has 8 bit resolution and a sampling rate of 1 GSa/s.
The transferability of the results to inexpensive, low-power~\gls{iot} hardware is discussed in Section~\ref{sec:discussion}.
The sensors are listed in Table I and are all available in the TO-92 packaging.
The frequencies are taken from the range $[10^3 Hz, 10^6 Hz]$ in 1 kHz steps, i.e. $f_{in} \in \{1000, 2000, \dots, 10^6 Hz\}$.
In this paper, we limit our considerations to computing the~\gls{rms} (also called~\emph{effective voltage}) of the recorded output signal.
The~\gls{rms} is an equivalent voltage which represents the DC voltage value that would produce the same heating effect, or power dissipation, in the circuit, as the applied AC voltage.\\

\begin{table}[h]
\caption{Sensor hardware used in our experiments.}
\label{tab:sensors}
\begin{tabular}{l|rrrr}
 & \textbf{TMP36} & \textbf{LM61} & \textbf{MCP9700} & \textbf{LMT85} \\ \hline
\rule{0pt}{2ex}
Min/max temp. (\degree C) & -40/+125 & -30/+100 & -40/+150 & -50/+150 \\
Accuracy (\degree C) & $\pm$ 2 & $\pm$ 2 & $\pm$ 1 & $\pm$ 0.7 \\
Min/max supply (V) & 2.7/5.5 & 2.7/10 & 2.3/5.5 & 1.8/5.5 \\
Supply current ($\mathrm{\mu}$A) & 50 & 125 & 6 & 5.4 \\ 
Manufacturer & AD & TI & Microchip & TI \\ \hline 
\end{tabular}
\end{table}

Our experiments shall give an insight on the granularity of identification that can be achieved using the proposed method:
Inter-device identification aims to distinguish between two different circuit layouts, whereas intra-device identification relies on differences between identical circuit layouts due to process variation.
The latter is of course the more difficult task, requiring more precise equipment and being less likely to succeed.

\subsection{Full Fingerprints}

\begin{figure} 
    \centering
  \subfloat[\label{fig:models_rms}]{%
       \includegraphics[width=0.49\linewidth]{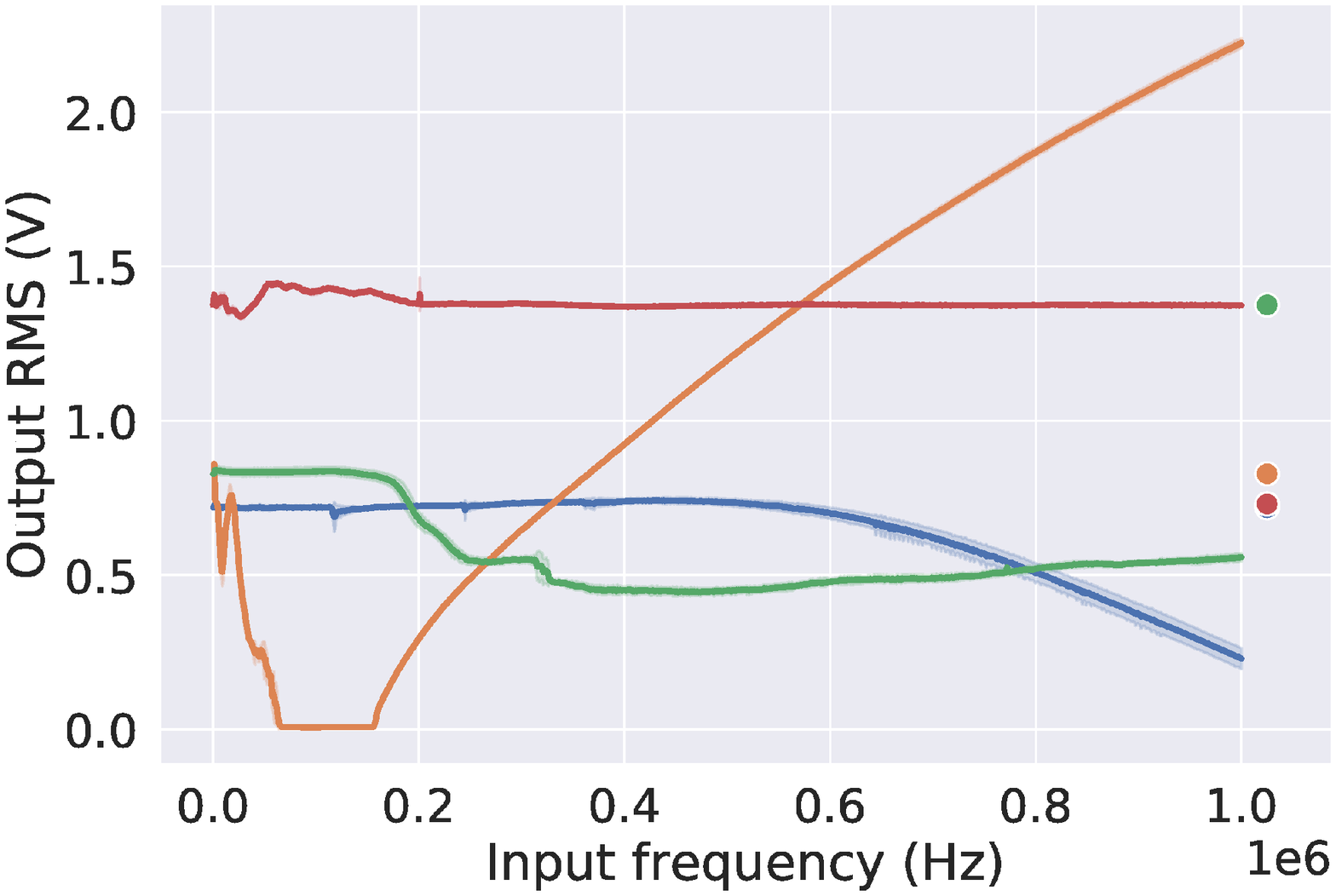}}
    \hfill
  \subfloat[\label{fig:models_var}]{%
        \includegraphics[width=0.49\linewidth]{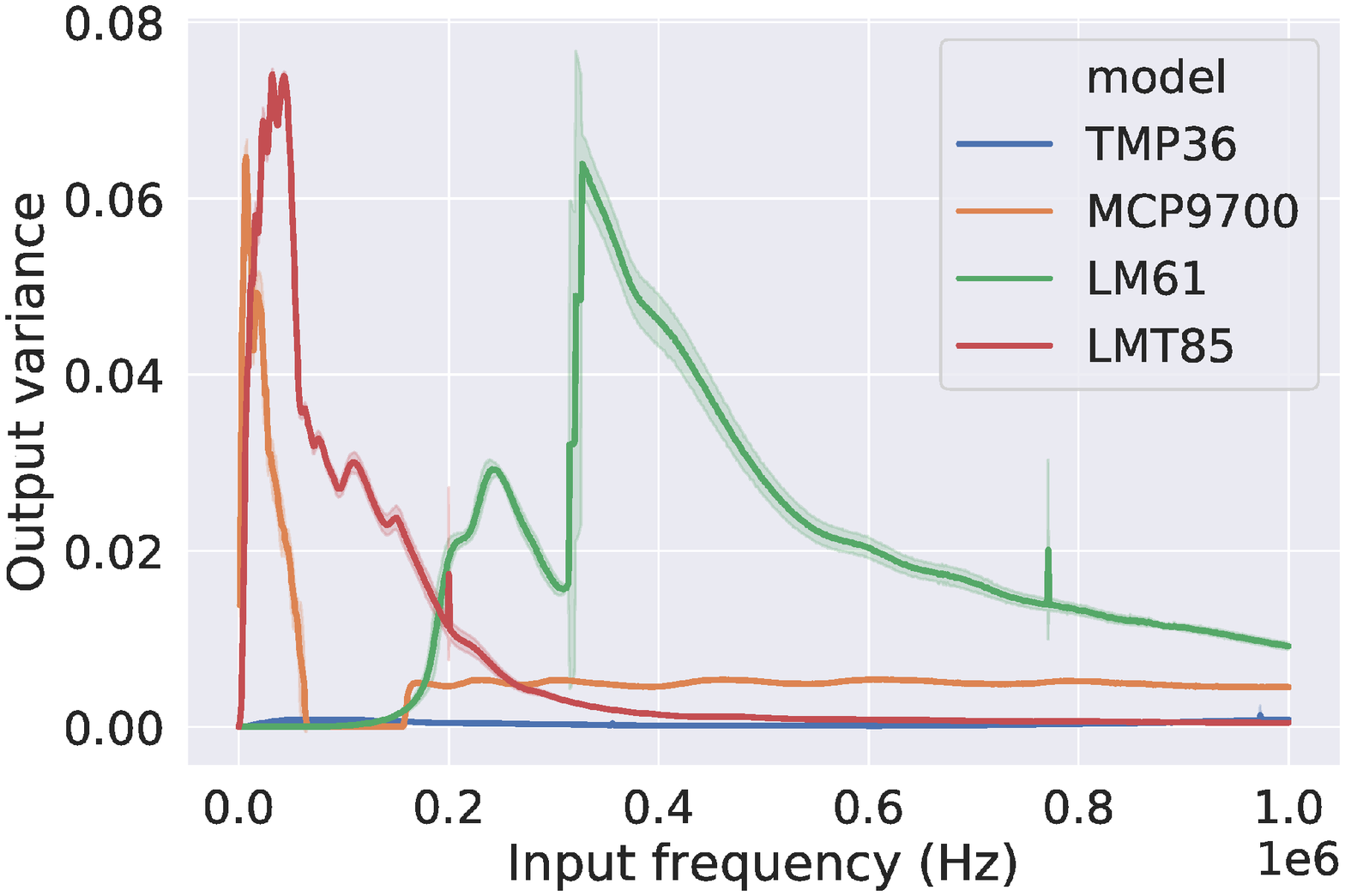}}
  \caption{(a) Output RMS per sensor model. Lines represent average RMS, shaded areas represent confidence intervals ($2\sigma$). The small dots at the right hand side mark the output voltage of each sensor model in normal operation mode and at room temperature. (b) Output variance per sensor model.}
  \label{fig:models_rms_var} 
\end{figure}

First, we qualitatively compare full fingerprints between different sensor models over the entire 1 MHz band.
Figure~\ref{fig:models_rms_var} displays~\gls{rms} and variance per 1 kHz step for the four tested models.
Solid lines represent the average value over three independent measurements with approximately the same environmental temperature ($\delta_T \approx .5^{\circ}C$).
Shaded areas around the curves show the standard deviation $\sigma$.
The dots in Figure~\ref{fig:models_rms} mark the measured output voltage of the sensors when supplied with 3.3V DC at room temperature.
These values are identical for TMP36 and LMT85 so their points coincide.
Visual inspection suggests that with the exception of a few crossing points, the differences between device models are large enough for reliable identification.
Interestingly, the variance of TMP36 never exceeds $10^{-6}$, indicating exponentially increasing dynamic impedance.
We also observe MCP9700 to entirely shut down for a certain band of frequencies around 100 kHz.
This could be due to a high reactance but further analysis is needed since the datasheets are inconclusive in this regard. 


\begin{figure}[htb]
  \centering
  \includegraphics[width=0.98\columnwidth]{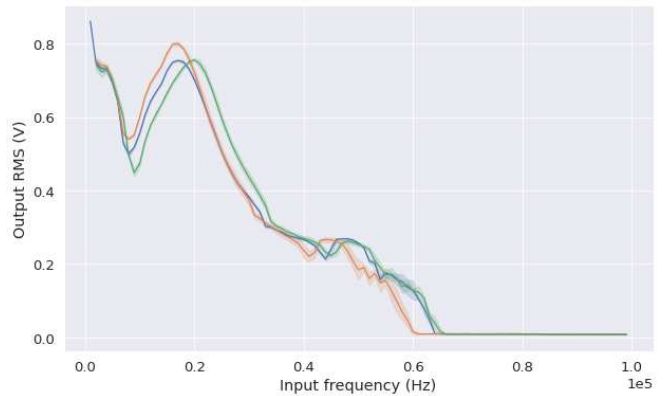}
  \caption{Output RMS per sensor instance in the range up to 100 kHz (MCP9700).}\label{fig:mcp_rms}
\end{figure}

For intra-device variation, we take an exemplary look at the output~\gls{rms} of three different instances of the MCP9700 sensor in Figure~\ref{fig:mcp_rms}.
The input domain (x axis) is narrowed to the first 100 kHz for increased readability. 
Again, solid lines correspond to mean values while shaded areas visualize the standard deviation w.r.t. multiple measurements of each device.
Visual inspection reveals differences beyond the confidence intervals but quantitative analysis is needed before conclusions can be drawn.

\subsection{Sensor Identification}

\begin{figure} 
    \centering
  \subfloat[\label{fig:points_rmse}]{%
       \includegraphics[width=0.49\linewidth]{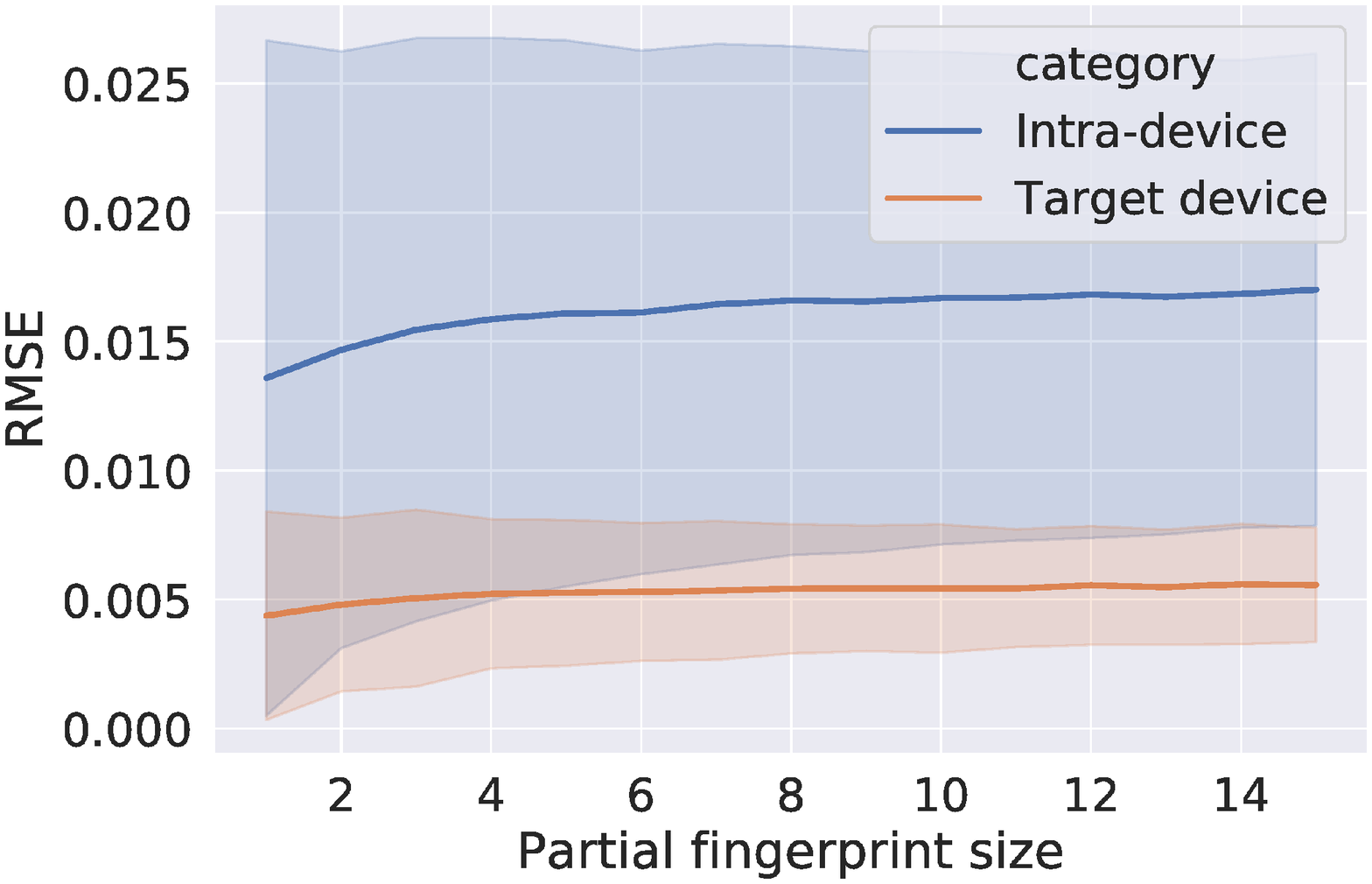}}
    \hfill
  \subfloat[\label{fig:f1_10points}]{%
        \includegraphics[width=0.49\linewidth]{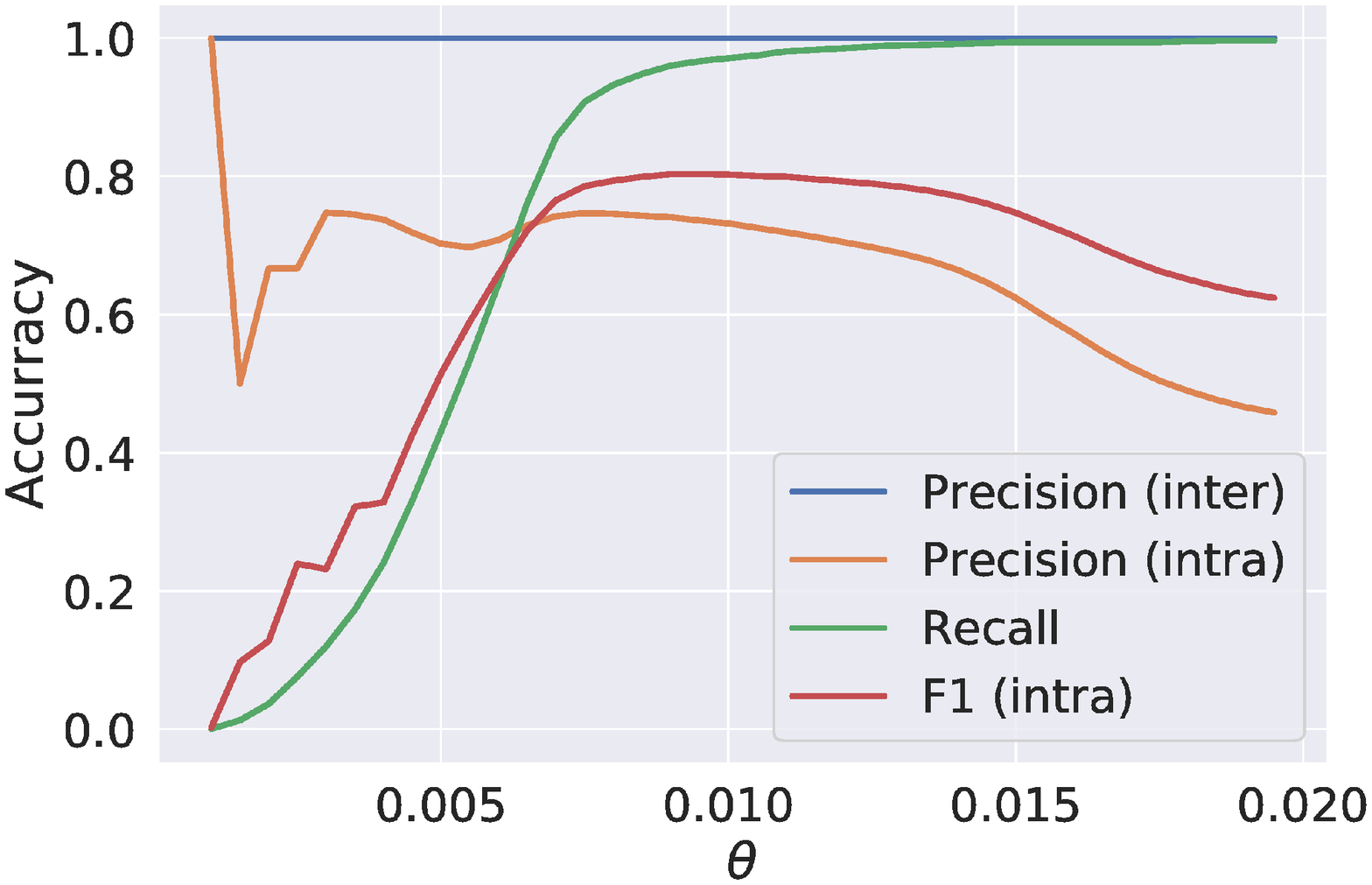}}
  \caption{Evaluation of the two hyperparameters. (a) RMSE for different sizes of the partial fingerprint. (b) Precision, recall and F1 score for various settings of $\theta$.}
  \label{fig:theta_P} 
\end{figure}

Following the authentication scheme presented above, we recall that a partial fingerprint is constructed from the~\gls{rms} responses to a set of challenge frequencies.
Authentication then proceeds by computing the deviation of the partial fingerprint from the previously obtained and stored full fingerprint.
The~\gls{rmse}, or identification error, is used to quantify the difference between a target partial fingerprint and the reference full fingerprint.
For a given full fingerprint $\hat{X}$ consisting of responses $\hat{x}_i, i=1,\dots,|\hat{X}|$ and a partial fingerprint $X$ with $|X| = P << |\hat{X}|$, we compute the~\gls{rmse} as follows:
\begin{equation}\label{eq:rmse}
    \epsilon \coloneqq \sqrt{\frac{1}{P} \sum_{p=1}^P (\hat{x}_p - x_p)^2}
\end{equation}
In~\gls{puf}-based authentication, usually the hamming distance is used, because responses consist of bit sequences instead of a set of real numbers.
It is reasonable to assume that the identification error decreases with an increasing number of points per partial fingerprint.
We assess this assumption by showing mean~\gls{rmse} values for different partial fingerprint sizes $|X|$ in in Figure~\ref{fig:points_rmse}.
In total, 1000 matchings between randomly chosen partial fingerprints and the corresponding full fingerprint were performed per device and partial fingerprint size.
We only compare~\gls{rmse} of target device matching with that of intra-device matching because the inter-device~\gls{rmse} is orders of magnitude higher and would distort the graph significantly. 
The graph confirms that with larger number of points per partial fingerprint, the difference between intra-device~\gls{rmse} and target device~\gls{rmse} increases and thus the identification error decreases.
Additionally, we observe the standard deviation of~\gls{rmse} to be nearly constant across fingerprint sizes for matchings with the target device itself, whereas it decreases for intra-device comparisons.
While the general implication seems to be ``the more points per partial fingerprint, the better'', we have to keep in mind that this parameter is also subject to practical constraints, since larger fingerprints require more power to record and communicate.


Once the~\gls{rmse} between partial and full fingerprint has been calculated for the target device, a decision needs to be made whether the device under scrutiny really is the expected one.
A straightforward approach is to define a threshold parameter $\theta$ and declare a given partial fingerprint to not match the corresponding full fingerprint if the~\gls{rmse} lies above that threshold.
In Figure~\ref{fig:f1_10points}, we investigate the relationship between fingerprinting accuracy and the threshold parameter $\theta$.
Accuracy is operationalized as F1 score which is calculated from the classical information retrieval concepts~\emph{precision} and~\emph{recall}.
The resulting curves for $\theta \in [0.001, 0.02]$ and 10 points per partial fingerprint are displayed in Figure~\ref{fig:points_rmse}.
As expected, recall improves for increasing $\theta$ up to a certain point because higher values correspond to a larger tolerance for deviations between partial and full fingerprint, e.g. due to noise.
With higher thresholds, the system would also accept fingerprints that differ due to process variation, resulting in an increasing number of intra-device false positives.
The best setting for $\theta$ under the tested conditions seems to be around $\theta = 0.01$ where the F1 score is maximal.\\

Finally, in Table II, we report the number of false positives (FP) and false negatives (FN) per device and granularity for a fixed partial fingerprint size of 10 and $\theta = 0.01$.
All numbers are given with respect to 1000 matchings.
The results show that there is, on average, a 2.9\% chance of a false mismatch with a fingerprint of the target device (false negative) over all four models.
This residual error is probably connected to small fluctuations of the environmental temperature, which is further discussed in the next section.
Just as with~\gls{puf} authentication, in such a case, another authentication attempt could be initiated, reducing the probability for a false negative to $0.084\%$.
As for false positives, the results indicate we would never mistake another model for the target device, but multiple instances of the same model can be misclassified as false positives.
This effect is less severe for some models, e.g. LM61 and very severe for others, e.g. LMT85.
Looking again at Figure~\ref{fig:models_rms_var} we intuitively understand why this is the case: The full fingerprint of LMT85 has nearly constant~\gls{rms} and very low variance for input frequencies above 200 kHz.
Hence for this sensor, in practice, we should constrain the fingerprinting to regions of the spectrum, where intra-device variance of the target device is high in order to achieve good results.

\begin{table}[]
\caption{Identification accuracy.}
\label{tab:fpfn}
\centering
\begin{tabular}{l|rrr}
        & \multicolumn{1}{l}{FN} & \multicolumn{1}{l}{FP (inter)} & \multicolumn{1}{l}{FP (intra)} \\ \hline
TMP36   & 6.5\%                  & 0\%                            & 1.7\%                          \\
MCP9700 & 2.2\%                  & 0\%                            & 29.9\%                         \\
LM61    & 1.3\%                  & 0\%                            & 0\%                            \\
LMT85   & 1.5\%                  & 0\%                            & 94.3\%                         \\
\hline
\end{tabular}
\end{table}




\section{Discussion of Practical Feasibility}\label{sec:discussion}

Our results show that analog fingerprinting is a suitable method for both inter- and intra-device identification.
As indicated before, a few open questions have to be addressed before the approach can be used in critical~\gls{iot} applications.
First, appropriate electrical components for generating the challenges (\gls{dac}) and reading the responses (\gls{adc}) have to be identified and tested.
The component properties must be balanced with respect to resolution, price and power consumption in order to be viable for large-scale deployments.
Currently, there are a number of candidate devices for such purposes on the market which will be tested in the next phase of our research.
In the same step, the accuracy of the approach could be improved by measuring the jitter curves of the~\gls{adc} and~\gls{dac} to account for their disturbance in the fingerprints.
Actually, we should account for the filter between~\gls{dac} and sensor as well, which also has a significant influence on the quality of the input signal and thus on the reliability of the fingerprint.
It might also be possible to use chips as used for~\gls{is} which contain high precision~\gls{dac},~\gls{adc}, and some additional circuits to compute high-level features such as~\gls{dft}.
Likewise, the~\gls{adc} could be replaced with an RMS-to-DC converter circuit to avoid recording multiple samples for manual RMS computation and thus save power.\\


Another critical question for the success of our method concerns the sensitivity of the device to changing environmental conditions.
That is, does the fingerprint change linearly with respect to the environmental temperature?
Many~\gls{iot} temperature sensors, including the four devices under scrutiny in this paper, provide so-called~\emph{linear slope} characteristics within the specified operation temperature range\footnote{cf. e.g. \url{https://www.analog.com/media/en/technical-documentation/data-sheets/TMP35_36_37.pdf}}.
The linearity is intended to allow for a first-order transfer function between measured voltage and sensed environmental temperature which simplifies device integration.
The reason why this is important for us is that during bootstrapping of the full fingerprints, the ambient temperature is very likely to differ from when the partial fingerprint is recorded.
Recall that in our experiments, this difference was controlled within $0.5^{\circ}C$.
However, if the effects on the fingerprint are linear, we can use efficient convex optimization methods to obtain a reliable fingerprint and at the same time determine the ambient temperature from the shift between partial and full fingerprint.
In other words, the fingerprint would implicitly contain the target observable, thus effectively fusing measurement with authentication.
Another solution would be to record multiple full fingerprints for different temperatures and use big data analysis methods to verify an incoming partial fingerprint.
Finally,~\gls{iot} applications in critical infrastructures usually ensure fault tolerance through sensor redundancy, which could be exploited for cross-verification of individual fingerprints between co-located nodes.\\

\section{Conclusion}\label{sec:conclusion}

In this paper, we presented a new method for fingerprinting sensors connected to an~\gls{iot} node for secret-free authentication.
The approach proceeds by challenging the device with a fixed-frequency alternating current, while observing the effective voltage at its outputs as the response.
We tested four commercially available~\gls{iot} temperature sensors using~\gls{rmse} error to compute pairwise similarities between fingerprints.
Our results show that the approach is, with bounded error, suitable for both inter- and intra-device identification.
We further elaborated on the effects of two hyperparameters contained in our method, the size of a partial fingerprint and the threshold for the matching error.
While the former is subject to a tradeoff between power consumption and accuracy, the latter balances false positives and false negatives.
Next steps with our approach include the implementation using low-power~\gls{iot} hardware and the integration of varying environmental conditions into our fingerprint model.

\bibliographystyle{IEEEtran}
\bibliography{refs}

\end{document}